\documentclass[aps,prl,twocolumn,groupedaddress]{revtex4}
\usepackage{amsmath}
\usepackage{graphicx}

\newcommand{\be}{\begin{equation}}
\newcommand{\ee}{\end{equation}}
\newcommand{\bea}{\begin{eqnarray}}
\newcommand{\eea}{\end{eqnarray}}
\newcommand{\ba}[1]{\begin{array}{*{#1}{c}}}
\newcommand{\ea}{\end{array}}

\newcommand{\pr}{{\rm Prob}}

\newcommand{\va}{v}

\begin{document}

\title{Spatial structures and dynamics of kinetically constrained models of glasses }

\author{Cristina Toninelli}
\email[]{cristina.toninelli@roma1.infn.it}
\affiliation{Dip. Fisica, Universita' La Sapienza, 00165 P.le A. Moro 5, Roma}

\author{Giulio Biroli}
\email[]{biroli@spht.saclay.cea.fr}
\affiliation{Service de Physique Th{\'e}orique, CEA/Saclay-Orme des Merisiers,
F-91191 Gif-sur-Yvette Cedex, FRANCE }

\author{Daniel S. Fisher}
\email[]{fisher@physics.harvard.edu}
\affiliation{Lyman Laboratory of Physics, Harvard University, Cambridge, MA 02138, USA}

\begin{abstract}
Kob and Andersen's simple lattice models for the dynamics of structural
glasses are analyzed.  Although the particles have only  hard core
interactions, the imposed constraint that they cannot move if surrounded by
too many others causes slow dynamics.
On Bethe
lattices a dynamical transition to a partially frozen phase occurs. In
finite dimensions there exist rare mobile elements that destroy the
transition. At low vacancy density, $v$, the spacing,
$\Xi$,  between mobile elements diverges exponentially or faster
in $1/v$. Within the mobile elements, the dynamics is intrinsically cooperative and the characteristic time scale diverges faster than any power of $1/v$ (although
slower than $\Xi$).  The  tagged-particle diffusion coefficient vanishes roughly as $\Xi^{-d}$. 

\end{abstract}


\maketitle

In many liquids dramatic slowing down occurs on cooling, equilibrium
cannot be achieved, and structural relaxation  becomes complicated and spatially heterogeneous
\cite{DeBenedettiStillinger,ErdigerWeitz,Glotzer}. Despite a
great deal of effort, these remarkable phenomena associated with the
 ``glass transition"  are still far from understood.  Indeed,  the most basic issues are unresolved: Is the rapid slowing down due to proximity to an equilibrium phase transition, albeit in a restricted part of phase space? Or is the underlying cause entirely dynamical? In either case, are there characteristic length scales that grow substantially near the glass transition? \\
Theoretical developments have been hampered by a shortage of models that capture essential features yet are simple enough to be analyzed.  Exceptions are {\it kinetically constrained} models  \cite{Jackle,ReviewKLG} based on
 the Ansatz 
  that the glass transition is caused by   geometrical constraints on
dynamical rearrangements with static correlations playing no role
(see e.g. \cite{ChandlerGarrahanKLG}).
These are also models for
granular media (see e.g. \cite{Sellitto})
for which slow dynamics occurs already at densities well below close-packing \cite{Jaeger}.\\ 
We  focus on one of the simplest kinetically constrained models, the Kob-Andersen model (KA) \cite{KA}. KA is a single component lattice gas  with no static interactions other than hard core exclusion and dynamics given by a continuous time stochastic process. Each particle attempts, at a fixed rate, to move  to a randomly chosen empty neighboring site; {\it but} the jump
 is allowed  only  if  both before and after the move the particle has 
no more than some number, $m$, neighboring particles.  
This corresponds to {\it vacancies} moving  only if the initial and final
sites have at least $s=z-m-1$ neighboring vacancies, with $z$ the
coordination number of the lattice. 
Since this dynamics satisfies detailed
balance, the 
trivial distribution that is uniform over all configurations with a fixed number of particles is stationary: 
there can thus be no equilibrium transition. 
Nevertheless, the dynamics is
sluggish and heterogeneous at high density \cite{KA,FranzKA}. 
For a three dimensional cubic-lattice with $s=2$,
fits of the self diffusion coefficient of a tagged-particle, $D_S$, strongly suggest a dynamical
glass
transition 
at $\rho_c\simeq 0.881$, above which $D_S$
appears to vanish and the structural relaxation time to diverge
\cite{KA}.\\
Our analysis shows that there are four classes of behavior for general KA models. These can be  characterized by the dependence of $D_S$ on the {\it vacancy density}, $v$: ({\cal N}) {\it normal} with $D_S\sim \va^q$ as $\va\to 0$; ({\cal C}) 
{\it collective freezing} with $D_S\to 0$ faster than any power of
$\va$ as $\va\to 0$ ; ({\cal T}) {\it dynamical transition} with $D_S\to 0$ as $\va \searrow \va_c$ with a non-zero critical $\va_c=1-\rho_c$; and ({\cal F}) {\it always frozen} with a finite fraction of the particles never moving for any $v$.
{\it Normal} behavior occurs if a finite cluster of $q$ vacancies can move through an otherwise-totally filled lattice,  e.g. a triangular lattice with $s=1$ in which vacancy pairs ($q=2$) are mobile. This  yields $D_S$ proportional to the cluster concentration, $v^q$. At the opposite extreme, a square lattice with $s=2$ is {\it always frozen} as a  four particle square can never move.
We focus on the  interesting intermediate cases,  {\cal C} and {\cal T}.
We show that a
{\it dynamical transition} 
takes place on (tree-like) Bethe lattices --- even with finite size loops.  But such a transition {\it cannot}
 occur on  finite dimensional lattices:  if neither normal nor always frozen, these exhibit {\it collective freezing} with the dominant dynamical processes involving a number of vacancies that diverges as $v\to 0$. This class ({\cal C}) includes the original $s=2$ cubic lattice.

We first analyze  Bethe lattices: infinite tree-like graphs with fixed connectivity $z$
which crudely approximate  high--dimensional or high--coordination--number lattices.
Monte-Carlo simulations 
of $10^{4}$ sites 
with $z=4$ and $s=1$ 
(crudely mimicking the $s=1$ square lattice) 
suggest a dynamical correlation length (as defined in \cite{FranzDH}) that diverges at a 
critical density $\rho_{c}\simeq 0.89$. \cite{long} 
Approaching this apparent dynamical transition the local--density correlation function displays two-step relaxation, as for supercooled liquids.  \cite{ReviewOE}

The tree-like structure of  Bethe lattices enables
analytic study by iteration.  Arranging the  tree with $k=z-1$ branches going
up from each node and one going down, we focus on the following 
events for a chosen {\it node} and rearrangements {\it restricted} to  the {\it subtree above it} : 
(${\cal A}$) node is occupied by a particle which cannot move up
even if the site 
below it is empty; (${\cal B}$) node is empty but a particle can never move onto it from below; (${\cal C}$)
node is occupied by a particle which can only move up if  the node below it is empty; (${\cal D}$) none of the above. Self-consistent equations for the probabilities of these events can be found by combining $k$ branches together with  the site
below them to which they are all linked \cite{long}. 
For $s=1$ with $z=k+1=4$,
this yields a critical density,
$\rho_c \cong 0.888825$ in  agreement with simulations.
 Below $\rho_{c}$, $\pr[{\cal A}]=\pr[{\cal B}]=0$;  particles are mobile and diffusive at long times.
At $\rho_c$, $\pr[{\cal A}]=A_c\neq 0$ and the fraction of immobile particles jumps discontinuously; so does the  {\it extensive configurational entropy},
$S_C=\ln[\# \mbox{different ergodic components}]$. \cite{long}  
Above $\rho_c$, $\pr[{\cal A}]-A_c \sim \sqrt{\rho-\rho_c}$, the singularity indicative of diverging length and time scales. 
The KA dynamical transition on this Bethe lattice (like $1/r^2$ percolation in one-dimension \cite{lr-perc})
thus has both critical and ``first order" characteristics.
Interestingly,
infinite-range {\it quenched random} p-spin models, conjectured to be related
to the glass transition, also exhibit
 a dynamical transition with similar 
features \cite{ReviewOE}.

The basic features found above hold  for
generic KA models on Bethe lattices with $s\neq 0,k$, 
including those with {\it finite size loops} 
\cite{long}, i.e. for the so called pure Husimi trees \cite{Monroe}
and for Bethe lattices where single sites are replaced by $L\times L$
squares (these interpolate among the Bethe lattice, $L=1$, and the 
square lattice, $L=\infty$).
The only exceptions are cases in which particles with two
occupied neighbors cannot move ($m=1$): the transition
is then continuous and $\rho_c=1/k$, the critical density for conventional
site percolation. This equivalence is due to the fact that the dynamical
transition is associated with the emergence of an infinite cluster of
permanently blocked particles. In general a frozen phase will only exist
if an infinite cluster remains after all the particles mobile under KA rules have been iteratively removed. This is
a type of {\it bootstrap percolation}, already introduced
in this \cite{KA} and broader \cite{Adler, Leath,AizenmanLebowitz}
contexts. For Bethe lattices,  an exact solution 
\cite{long} (see \cite{Leath} for a simpler example)
yields the same critical density found above.

We now turn to finite dimensions  and show that the dynamical transition is destroyed by exponentially rare
processes which will only occur in sufficiently large systems. Nevertheless, at high densities  the dynamics is intrinsically collective and extremely slow. Furthermore,  the ghost of the Bethe lattice transition may cause a sharp crossover at an apparent critical density. 

 We focus initially  on the simple $s=1$ square lattice case. 
First, let us define a configuration as {\it framed} if all its boundary sites are
empty (see for example the square on the right of fig. 1); and as {\it
frameable} if, by an allowed sequence of moves,  a {\it framed}
configuration can be reached (see for example the square on the left of fig. 1). 
A key observation is that the lines of vacancies on the boundaries of a framed square can be
 shifted (moving vacancies starting from the corner) so
that any nearest neighbor pair of particles inside the square can be sandwiched
between two lines of vacancies. 
The
position of this pair of particles can then be exchanged 
and the vacancy lines afterwards shifted back to their original
positions. The net result is a pair exchange of particles
with all other particles returned to their original positions.
By combining such pair exchanges all  frameable configurations with
the same density can be connected. Thus, these form an  irreducible
component $\cal Z_\rho$ of the configuration space. 
We must now show that almost all random configurations with a given $\rho<1$ belong to $\cal Z_\rho$ in the
thermodynamic limit. 
This can be  done iteratively by observing that an $l$ by $l$ frameable
configuration that has at least two vacancies externally adjacent to each of its sides is
also an $l+2$ by $l+2$ frameable configuration, see fig. 1. Starting from a two by two frameable
``nucleus" of vacancies, one can thus grow an $L$ by $L$
frameable configuration if the requisite vacancies are present in each concentric shell. The probability that this occurs is 
$P_{L} (\rho )=(1-\rho )^{4}\prod_{l=2}^{L/2}
[1-\rho ^{2l}-2l\rho ^{2l-1}(1-\rho)]^{4}$
which converges to a non-zero  probability,
$P_\infty(\rho) $. Thus for all $\rho$ infinite frameable squares exist with all frameable configurations within them being reachable. For small vacancy densities, $P_\infty(\rho) \simeq
e^{-2 \hat{K}_F/v} $
 with $\hat{K}_{F}\simeq 4.48$ .
Moreover,  $P_L(\rho)$ depends weakly on
$L$  for $L>\xi(\rho) \sim\ln (1/\va )/\va$; $\xi$ is
thus the {\it core} size of frameable regions. Note that $P_L$ is the probability that a frameable square can be constructed around a nucleus at a {\it fixed} position.
On the other hand the probability, $F_L(\rho)$, that a large square
is frameable is roughly given by considering  the $O(\left[L/\xi\right]^2)$ possible positions of a nucleus within the square. Although the probabilities of a configuration being frameable around each of these possible nuclei are not independent, if the nuclei are separated by more than a core diameter they are roughly independent. Indeed, it can be
proven that $F_L$ asymptotes to one for $L\gg \Xi$ with the long crossover length scale $\Xi(\rho)$ given, as expected from the above argument,  by
$\Xi^2 P_\xi(\rho)\sim \Xi^2 P_\infty(\rho) \sim \xi^{2}$.

We have shown that at any density almost all configurations of sufficiently large systems can be reached from one another. This   {\it excludes} the possibility of a dynamical transition  \cite{footnoteergodicity}.
 The phase space of systems with linear size, $L$, larger than the crossover length $\Xi$  is covered almost entirely by a single ergodic
component, while the phase space for $L<\Xi$  is typically decomposed into many disjoint parts. The framing analysis yields an {\it upper
bound} for $\Xi$.

A lower bound for $\Xi$ can be obtained from
bootstrap percolation arguments.
>From a random configuration with density $\rho $ iteratively
remove all particles with less than three occupied neighbors. If
any particles remain they  must form a system-spanning cluster of frozen
particles and  the configuration space will thus be broken up 
into many disjoint pieces. This happens with high probability for $\ln L< K_F/\va$ with $K_F=\pi^{2}/18$ \cite{AizenmanLebowitz,new-bootstrap},
 yielding a lower bound for $\Xi$ with the same density dependence as the above upper bound, but a different constant $K_F$ less than $\hat{K}_F$ of the upper bound.
However,  a better definition of
framed configurations yields a  $\hat{K}_F$  which is identical to
$K_{F}$ \cite{long}.
The corresponding frame is {\it minimal} and
the asymptotic form of the crossover length  $\Xi \sim
e^{\pi^2/18\va}$ {\it exact} (up to subdominant factors).

\begin{figure}[bt]
\centerline{\includegraphics[width=0.9\columnwidth]{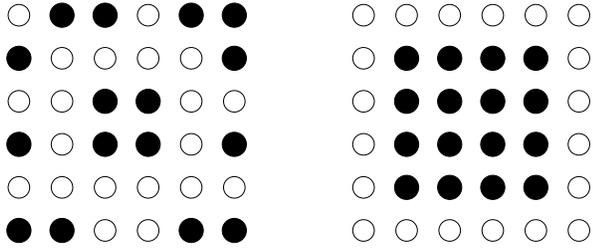}}
\caption{
On the left there is a $4$ by $4$ framed configuration with two vacancies (white circles) adjacent
to each side. After moving the external vacancies at the
corners the framed square can be expanded to the $6$ by $6$ framed
configuration at right.}
\label{fig}
\end{figure}

All other finite-dimensional KA models can be analyzed by generalizing the notion of frameable configurations, we thereby
rule out {\it any}  dynamical transitions. For $d$-dimensional hyper-cubic lattices  $s=0$ and $s>d-1$ correspond respectively to {\cal N} and {\cal F} while all the non-trivial cases, $1\le  s \le
d-1$, are class {\cal C}, collective freezing. \cite{long}
Analogous expansion arguments yield upper bounds on their crossover lengths:
$\Xi< exp^{\circ s} \left( K_U (s,d)/v^{\frac{1}{d-s}}\right)$ with $exp^{\circ s}$  the
exponential function iterated $s$ times and $K_{U}(s,d)$ a positive
constant given in \cite{long}. Bootstrap percolation yields  a lower bound of the same form
but with a different constant $K_{L} (s,d)$ \cite{cerfcirillo}
The scale, $\xi$, above which the probability that a nucleus is expandable
saturates, is generally of order $\ln \Xi$, up to log-log factors.

We now turn to the dynamics,
focusing again on the $s=1$ square lattice model at high density.
Vacancies will typically be far
apart or in small clusters that cannot move. However vacancies can move within the cores  of large
frameable regions. And these cores of size $\xi \sim  \ln (1/v)/ v$ are themselves mobile because they are likely to find vacancies on {\it all} the $\xi$ successive line segments needed for them
to move in a given direction.
A generic particle cannot move substantial distances except when a
mobile core passes by; assuming independent core motion a tagged particle will thus diffuse with
$D_S\sim D_Mn_M$, where $n_M\sim 1/\Xi^2$ is the density of the mobile cores
 which diffuse with $D_{M}\simeq
\xi^{2}/\tau_\xi$, $1/\tau_\xi$ being the typical relaxation rate of a core. Using our framing analysis and a generalization of the
technique of \cite{Spohn}, we have proved that, indeed, $D_S>0$ for all $\rho$. \cite{long}

The
relaxation time,  $\tau_\xi $, of a core is dominated by the time to reach the most severe {\it bottleneck}
in configuration space. Since all  configurations are equiprobable
this is proportional to $\exp (\Delta S)$, 
 the ratio of the  number of accessible (frameable)
configurations of the core to the number in the bottleneck \cite{long}.
The worst case scenario in which there is a {\it single} bottleneck configuration  corresponds to $\Delta S=S_{\rm total}\simeq \ln
\xi!$, yielding an upper bound $\tau_\xi < \xi!$.
A lower bound is obtained by noting \cite{long} that to equilibrate a frameable square of size $\xi$ one has to pass through
configurations with the nucleus in a {\it
corner} of the square. Using a transfer matrix technique we obtain for
$\ell\times\ell$ minimally frameable squares an entropy difference
$\Delta S\approx \Upsilon\sqrt{\ell} + \alpha \ln \ell + C$ with
$\Upsilon= 2\sqrt{6+\sqrt{22}}-2\sqrt{3}\cong 3.075$ and $\alpha$
computable in principle. \cite{long}  Simulations of frameable squares with $\ell=4$ to $\ell=16$ yield equilibration times with $\log \tau_{\ell}$ increasing
 slower than $\ell $ and in good agreement
 with the square root law with the {\it predicted} coefficient $\Upsilon$
plus logarithmic corrections. We thus conjecture that the  entropic bottlenecks for moving the core of a mobile region  are equivalent, up to subdominant factors, to
the entropy loss associated with forcing the nucleus of
of such core to a corner.  


As long as bottlenecks for motion of the mobile cores are not too much larger than the above estimates, $\tau_\xi\ll\Xi^2$ and 
the dominant contribution to $D_S\sim D_Mn_M$ will arise from the low density of mobile regions rather than the long time needed for motion within them.
Thus, to leading order in $\va^{-1}$, $\ln D_S\approx 2K_F/\va$
for the square lattice model with $s=1$.
Numerical simulations at high densities
fit this form well, see fig. 2: in particular
$\lim_{v\rightarrow 0}v\ln D=2K_F\cong
0.9-0.95$ {\it cf} the predicted $2K_F\cong 1.1$.
\begin{figure}[bt]
\centerline{
\includegraphics[width=0.9\columnwidth]{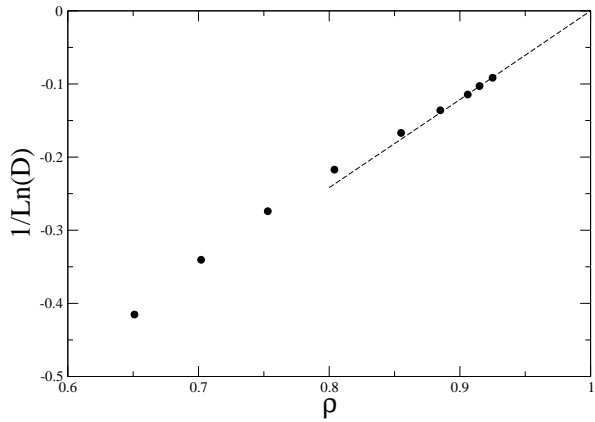}}
\caption{$(\ln D)^{-1}$  as a function of $\rho $ for
$400$ by $400$ square lattice with $s=1$. 
Straight line  shows  $\ln D_S\propto 1/v$ behavior.} 
\label{fig2}
\end{figure}
For the other KA models, we obtain similar relationship between $D_S$ and $n_M$.
For the $s=2$ cubic case theis yelds $\ln \ln D^{-1}_{S}\propto
1/v$ , a result consistent with a recent finite size scaling analysis \cite{Berthier}.

Thus far we have focused on the high density limit.
Yet simulations on KA and other kinetically constrained models suggest a transition at a non-trivial critical density \cite{KA}. This is a natural consequence of the existence of a dynamical transition on Bethe lattices: in finite dimensions, the transition will be  replaced by a crossover which, in some cases, could be quite sharp. In the $s=1$ square lattice case a three--vacancy element can move
along a network of other vacancies provided these are linked no more weakly than via second
neighbors (along axes or diagonals).
At low density, there will always be a percolating vacancy network of such type. As the density increases, diffusion slows down as the vacancy percolation  cluster, on which mobile elements rely, sparsens.
Beyond the second-neighbor vacancy percolation transition at $\rho_{2P}$,   the {\it simple} mobile elements can no longer move. But by {\it growing}, mobile elements can still form: they have to become large enough to find a percolating cluster on which objects of their size can move. At high densities, such mobile elements will be the cores of (minimal) frameable regions that we have
discussed above.
If the crossover from small mobile objects and behavior similar to that on Bethe lattices is sharp enough, a substantial range of  ``critical" behavior   of $D_S$ --- and of relaxation times --- might be observed near to an {\it apparent} transition at
$\rho_G\approx\rho_{2P}$.
This reasoning can be generalized to other  KA models --- indeed a sharper crossover should occur in higher dimensions --- and could be
an explanation of the apparent dynamical transition found for the
original cubic lattice KA \cite{KA}.

We have shown how simple kinetically constrained models exhibit concretely some of
the qualitative features conjectured to be the
cause of the dramatic slowing down in structural glasses.  In
particular, finite-dimensional models  can ``almost'' have a 
a dynamical phase transition; such exists on Bethe lattices and
 percolation arguments suggest  it is replaced by a
crossover --- possibly very sharp --- in finite dimensions. The dominant dynamical processes involve cooperative rearrangements of regions whose size diverges in the high density limit; nevertheless, these yield a non-vanishing
self-diffusion coefficient at all densities. But the 
length and time scales of these processes  grow extremely
rapidly at high densities. This is reminiscent of  the super-Arrhenius increase of the relaxation time as fragile liquids are cooled.\\

\begin{acknowledgments}
We thank L. Berthier, L. Bertini, J-P. Bouchaud, K. Dawson, J-M. Luck and
M. M{\'e}zard for interesting discussions.
CT thanks COFIN MIUR-2002027798 for financial support.
DSF thanks the National Science Foundation for support via DMR-9976621 and
DMR-0229243.
We thank Les Houches summer school LXXVII where this collaboration
initiated.
\end{acknowledgments}


\begin{thebibliography}{99}
\bibitem{footnoteergodicity}
 For the KA model one can also prove that irreducibility implies no ergodicity breaking in the thermodynamic limit (see \cite{long}).
\end{thebibliography}
\end{document}